\documentclass[aps,prl,reprint, twocolumn,showpacs,floatfix]{revtex4}
\usepackage{graphicx}

\begin{document}
\title{Photoluminescence Spectroscopy of many-body Effects  in Heavily Doped  Al$_x$Ga$_{1-x}$As}

\author
{Subhasis Ghosh and Niladri Sarkar}

\affiliation{School of Physical Sciences, Jawaharlal Nehru University, New Delhi 110067}

\begin{abstract}

We present an  experimental investigation of heavy doping-induced many-body effects such as   band gap
narrowing(BGN) and Fermi edge singularity(FES) in   Al$_x$Ga$_{1-x}$As using photoluminescence(PL)
spectroscopy.  The  band-to-band
transition energy shifts to lower energies  and FES-like feature in PL spectra grows  with increasing  electron
concentration. We show that FES-like feature is a nonmonotonic function of temperature and excitation intensity. Our data lead us to suggest that a small concentration of nonequilibrated holes is required to enhance the FES-like feature in the  PL spectra.
\end{abstract}

\maketitle

 There are two most important many-body effects related to heavy doping in III-V semiconductors, (i) the shift of conduction and valence band towards each other causing band gap narrowing(BGN) due to electron-electron interaction\cite{hk96}, and (ii) the many-body enhancement of the oscillator strength of the optical transition  near Fermi level causing  Fermi edge singularity(FES), due to the response of Fermi sea  to a photogenerated hole\cite{gdm90}.  The presence of the large concentration of free carriers can
cause a significant reduction of unperturbed band gap in
semiconductors due to  many-body interaction between electrons and holes and  carrier-impurity interaction\cite{sg00}.  Though a detailed experimental and theoretical analysis that rigorously combines the many-body effects due to exchange-correlation and statistical fluctuation  is extremely complicated problem, one can investigate the role of  many-body effects on BGN  in   alloy semiconductor by choosing the doping concentration carefully\cite{sg00}. 
FES is one of the most widely investigated  many-body effects, first theoretically predicted by Mahan\cite{gdm67}. This effect was first successfully understood in terms of threshold singularity in X-ray absorption and emission spectra in simple metals\cite{ko90}. In case of semiconductors, it was first observed\cite{msk87} in 2D electron gas in modulation doped InP/In$_x$Ga$_{1-x}$As quantum well and subsequently,  in several other systems based on low dimensional semiconductor structures(LDSSs).  It has been shown that the essential condition for observing the FES is the localization of the photogenerated holes. Typically the FES is identified  as an enhancement of  the oscillator strength of a transition close to Fermi level in PL spectra.   

Although it appears that FES is the simplest nontrivial many-body effect and understood experimentally and theoretically, upon closer inspection it seems several experimental results such as  (i) observation\cite{sab97,hpv99} of FES  in GaAs-based LDSSs without  substantial hole localization, (ii)  strong dependence\cite{hk00,iab97} of FES on carrier concentration and excitation intensity in GaAs- and InP-based LDSSs even in the presence of strong hole localization, and (iii) finally observation\cite{yim02,hk02} of FES at relatively higher temperature, are not understood within the existing theoretical framework. Moreover,   LDSSs have  been used to observe FES, but there are two inherent shortcomings in these systems,  (i) it is known that non-Coulumbian intersubband scatterings can induce FES as well\cite{wc92}, so  it is not known exactly the relative importance of  the extrinsic intersubband scattering and intrinsic many-body electron-hole interaction on the formation of FES\cite{tm00}  and (ii) the exact role of hole localization and reduced dimension and their relative importance on the formation of FES  is not completely understood in LDSSs.

In order to address these issues,  we have investigated  the formation of FES in higher dimensional systems, where the exact role of hole localization on FES may be understood unambiguously. If, the localization of photo-generated hole can be achieved, it should be possible to observe FES in degenerate bulk semiconductors\cite{aer87}.  A natural point to start, when studying heavy doping-induced many-body effect is BGN, which is probably the most well understood many-body effect and can be used to tune the range of electron concentration to observe the FES-like feature in PL spectra in heavily doped  Al$_x$Ga$_{1-x}$As. We show first the BGN in moderately heavy doped n-type Al$_x$Ga$_{1-x}$As is governed by many-body effects. This has been validated by us in a previous study\cite{sg00} using room temperature PL  and reconfirmed in this work by low temperature PL study.  This  is then  used to identify the FES-like feature in bulk Al$_x$Ga$_{1-x}$As. Emphasize is put on the anomalous temperature and excitation intensity dependence  of FES. Finally a scenario based on the exchange splitting of FES i.e. many-body excitonic level  has been invoked to model the nonmonotonic temperature dependence of  FES. 

Samples used in this investigation are high quality   Al$_x$Ga$_{1-x}$As epitaxial layers with     doping level ranging from 3$\times$10$^{16}$
to 3$\times$10$^{18}$cm$^{-3}$, grown on either  semi-insulating and
n-type GaAs substrate by   metal
organic chemical vapor epitaxy.  Se   is chosen as dopant  for it's  high solubility, low compensation    yielding highest doping level and minimum    DX-center related problems\cite{jcb90}.  The electron concentrations were determined  by Hall
measurements.  The active Al$_x$Ga$_{1-x}$As layers are
separated from the GaAs substrate by an undoped spacer layer to
avoid any 2D effect. Details about the samples are given in Ref.3.
  Room temperature and low temperature PL spectra of n-type Al$_x$Ga$_{1-x}$As as a function of electron concentration  for x =0.22 and x=0.11 are shown in Fig.1 and Fig.2, which  show the evolution of  low temperature PL spectra of Al$_{0.22}$Ga$_{0.78}$As and
Al$_{0.11}$Ga$_{0.89}$As as a function of electron concentration, respectively.  In case of  Al$_{0.22}$Ga$_{0.78}$As(Al$_{0.11}$Ga$_{0.89}$As),  the sample with lowest electron concentration of 3.5$\times$10$^{16}$cm$^{-3}$(5.4$\times$10$^{16}$cm$^{-3}$)  in sample 1A(2A), the dominant peaks at 1.842eV(1.689eV) is due to bound exciton(BE) and the broad peak at 1.822eV(1.672eV) is due to free-to-bound(FEB)- and donor-acceptor(DAP)-related transitions. A high energy shoulder at 1.847eV(1.695eV) is due to either band-to-band(BB) or free exciton(FE).  As the electron concentration is slightly increased to 4.5$\times$10$^{16}$cm$^{-3}$(8.2$\times$10$^{16}$cm$^{-3}$) in sample 1B(2B), almost similar PL spectra are observed in case of sample 1B(2B), but as the electron concentration increases to 1.3$\times$10$^{17}$cm$^{-3}$(2.4$\times$10$^{17}$cm$^{-3}$) in sample 1C(2C) BB  peak appears in lieu of BE and FE peaks due to  quenching of free excitons and screening of impurity potential. As the electron concentration is slightly increased to 3.1$\times$10$^{17}$cm$^{-3}$(3.7$\times$10$^{17}$cm$^{-3}$), almost similar PL spectra are observed in case of sample 1D(2D).   Now, as electron concentration is further increased 8.3$\times$10$^{17}$cm$^{-3}$(8.9$\times$10$^{17}$cm$^{-3}$) in sample 1E(2E), an extra high energy peak at 1.858eV in addition to BB, FEB and DAP peaks is observed pronouncedly in sample Al$_{0.22}$Ga$_{0.78}$As and as weak a shoulder  at about 1.724eV in sample Al$_{0.11}$Ga$_{0.89}$As.  Again, as the electron concentration is further increased to 9.6$\times$10$^{17}$cm$^{-3}$(2.1$\times$10$^{18}$cm$^{-3}$) slightly in sample 1F(2F), similar PL spectra as those in 1E(2E) is observed.  Energetic position and saturation behavior have revealed the identification of these peaks\cite{lp94}. Following are the general  features of the PL spectra in Al$_{0.22}$Ga$_{0.78}$As and Al$_{0.11}$Ga$_{0.89}$As.

(a) The transitions due to FE, BE, FEB and DAP are observed in samples with low electron concentration.   As the electron concentration is increased by order of magnitude, FE and BE peaks are replaced with BB transition, but FEB and DAP peaks are observed in all samples.

(b) Typical BB transitions are observed in room temperature PL spectra(Fig.1 and Fig.2), which  are characterized
with Maxwell-Boltzman lineshape\cite{sg00} in the high energy side and
relatively sharp cut-off in the low energy side caused by
(E$_G$-E)$^{1/2}$ dependence, where E$_G$ is the band gap.  It is clear that BB peaks  shifts towards lower energy as the electron concentration increases.

(c) The redshift of BB peak is also observed at  low temperature. Insets of Fig.1 and Fig.2 show the BGN as a function of  electron concentration.   

(d) The width of the all peaks, which are due to BB, FEB and DAP increases with electron concentration.

For determining the role of
many-body effects due to exchange and correlation, it is
desirable to restrict the doping concentration such a way that
typical BB lineshape can be observed in the PL spectra. It
has been shown that the effective change in the unperturbed band gap E$_G^0$ due to many-body
interactions  can be
given by\cite{gdm80}, $E_G= E_G^0-an^{\frac{1}{3}}-bn^{\frac{1}{4}}$, where n is electron concentration and first term is due to exchange interaction, which
comes from the spatial exclusion of the like spin away from each
other.  The second term is due to correlation energy, which
comes from the repulsion of the like charges, so that they do
not move independently but in such a way as to avoid each other
as far as possible. The values of the  coefficients a and b, which are  evaluated  by fitting the experimental data(as shown in insets of Fig1. and Fig.2) with above expression are 1.6$\times$10$^{-8}$ and 8.2$\times$10$^{-8}$  in Al$_{0.22}$Ga$_{0.78}$As and   8.2$\times$10$^{-9}$ and 2.1$\times$10$^{-8}$ in Al$_{0.11}$Ga$_{0.89}$As, respectively.     It is
not possible to make any comments on the value of coefficients
a and b either at 300K  or at 1.8K, because neither detailed theoretical,
nor experimental results are available on BGN in
Al$_x$Ga$_{1-x}$As.  Hence BGN at high  and as well as at low temperature governed by  intrinsic many-body interaction i.e. exchange and correlation among the carriers.

Following are the important features of the high energy peaks observed in samples 1E and 1F in Al$_{0.22}$Ga$_{0.78}$As and samples 2E and 2F in Al$_{0.22}$Ga$_{0.78}$As

(a) The extra high energy peak at 1.858eV in case of sample Al$_{0.22}$Ga$_{0.78}$As(1E and 1F) and a shoulder at 1.724eV in case of sample Al$_{0.11}$Ga$_{0.89}$As(2E and 2F) are not observed in previous PL studies on Al$_x$Ga$_{1-x}$As. The variation of the band-gap in Al$_x$Ga$_{1-x}$As due to alloy fluctuation determined  by scanning the sample with bound exciton
peak energy has been found to be   small ($\sim$0.4 meV/cm) across the sample and this fluctuation
corresponds the variation of $x$ less than 0.001 per cm and rules out
the possibility of energetic shift of BB or excitonic
transition towards higher energy due to band gap variations.

(b) The energetic position of this new peak is almost 20meV above the band gap. We have found that the position of Fermi level in samples with electron concentration of 8.3$\times$10$^{17}$cm$^{-3}$ to 9.6$\times$10$^{17}$cm$^{-3}$ in  Al$_{0.22}$Ga$_{0.78}$As  and 8.9$\times$10$^{17}$cm$^{-3}$ to 2.1$\times$10$^{18}$cm$^{-3}$ in Al$_{0.11}$Ga$_{0.89}$As is around 20meV at low temperature($<$10K).   

(c)  The high energy peak is completely quenched in  sample(1G)  with 2.8$\times$10$^{18}$cm$^{-3}$ electron concentration, but BB, FEB and DAP peaks are present, as shown in Fig.1.

(d) Fig.3 shows the evolution of PL spectra of sample 1F(Al$_{0.22}$Ga$_{0.78}$As), with excitation energy. At lowest intensity, the high energy peak is barely visible, but the  spectral feature  increases as excitation intensity increases. Further increase in excitation intensity smears out the peak.

(e) Fig.4 shows the temperature dependence of PL spectra for sample 1F(Al$_{0.22}$Ga$_{0.78}$As). The temperature dependence of integral PL intensities of BB, FEB and DAP peaks shows monotonic dependence, but the high energy peak shows completely opposite behavior. The integral intensity of this peak is observed to be a nonmonotonic function of temperature with a maximum at about 40K and this peak persists till higher temperature $\sim$70K. Similar nonmonotonic temperature dependence has also been observed(not shown) for high energy peak in  Al$_{0.11}$Ga$_{0.89}$As.     Non-monotonic temperature dependence of FES-like peak results S-shaped shift behavior of BB peak. The deconvoluted BB peak, as shown in inset of Fig.4, shows usual temperature dependence.

 All these findings indicate that  the high energy peak  is intrinsic in
nature. We attribute the origin of this peak  as due
to many-body effect {\it i.e.,} FES, which  gets further support from
Fig.1 and Fig.2. This transition   can be observed only in samples with  narrow range of electron concentration of 8.3$\times$10$^{17}$cm$^{-3}$ to 9.6$\times$10$^{17}$cm$^{-3}$ in  Al$_{0.22}$Ga$_{0.78}$As  and 8.9$\times$10$^{17}$cm$^{-3}$ to 2.1$\times$10$^{18}$cm$^{-3}$ in Al$_{0.11}$Ga$_{0.89}$As. This may be one of the reasons for not observing this FES-like feature in previous PL spectra of n-type Al$_x$Ga$_{1-x}$As.  Disappearance of this peak  in very heavily doped samples($>$3$\times$10$^{18}$cm$^{-3}$)
may be due to two reasons,  (i) it  has been shown\cite{ph92}
  that width of the luminescence spectra
due to FES increases and the intensity decreases with increasing
carrier density due to screening and movement of the electron layer from
the localized hole and     (ii) potential fluctuation due to the ionized donor can destroy
the FES by smearing out the sharp Fermi surface when the doping is more than 10$^{18}$ cm$^{-3}$.  This is indirectly corroborated by our results on BGN at low and high temperature on  the same samples. BGN is governed by mainly electron-electron interactions in those moderately doped samples(n$\leq$ 10$^{18}$ cm$^{-3}$) which show FES-like feature. We have observed that FES-like feature is completely quenched in samples with doping concentration more than 10$^{18}$cm$^{-3}$, which do not show typical BB lineshape.     Here we would like to
mention that weak appearance of FES in Al$_{0.11}$Ga$_{0.89}$As compared to that in Al$_{0.22}$Ga$_{0.78}$As may  be due to the fact that
alloy fluctuation in the dilute alloy is not enough to localize the
photogenerated holes efficiently. 
Localization of photogenerated holes
by alloy fluctuation plays an important role in the PL spectra of
moderately heavy doped degenerate bulk semiconductor relaxing the
{\bf k}-selection rule of photogenerated  holes and provides
the recombination channel with electrons occupying all {\bf k}-states upto
Fermi-edge.

Fig.3 and Fig.4 show the  temperature  and excitation intensity dependences of the PL
spectra of  sample 1F.  The observed nonmonotonic temperature and excitation intensity dependences of FES is not anticipated theoretically, because all previous theoretical investigations predict strong thermal quenching of FES due to smearing out of the Fermi surface. It is very clear that the nonmonotonic dependence of FES on temperature and excitation intensity implies that optimal temperature and excitation intensity are required for the observation of FES. Similar  dependences of FES has been observed\cite{yim02,hk02} in pseudomorphically strained modulation doped Al$_x$Ga$_{1-x}$As/In$_y$Ga$_{1-y}$As/GaAs  heterotructures.   Here, we present a scenario in which,  nonmonotonic temperature dependence of PL intensity(I$_ {PL}$) is modeled by two-level system. If we consider FES is more like localized many-body exciton, the nonmonotonic temperature dependence of I$_{PL}$ can be described by exchange splitting of localized excitononic level\cite{pdjc93}  by an energy $\Delta$ between threefold degenerate triplet state and the higher lying singlet state, with radiative rate of R$_T$ and R$_S$, respectively. At thermal equilibrium the total radiative decay R$_R$ can be given by $R_R=\frac{3R_T + R_S exp(-\Delta/k_BT)}{3 + exp(-\Delta/k_BT)}$. At low temperature only the lower lying triplet state is occupied and the radiative decay rate is small, because the optical transition from a pure triplet state is forbidden, but ,  spin-orbit interaction mixes  triplet and singlet states  and allows optical transition  with very low quantum yield. As the temperature increases, higher lying singlet state is occupied and quantum yield of the optical transition or the I$_{PL}$ of FES-like peak increases and attains a maximum, but, further increase in temperature leads  to thermal quenching of the bound states resulting the decrease of I$_{PL}$ resulting nonmonotonic temperature dependence of FES.   Inset of Fig.4 shows the fit to the experimental data with our phemenological model. We have obtained a value of 6meV for $\Delta$. We have found that the  $\Delta$ for exciton is close to 0.5meV in Al$_{0.22}$Ga$_{0.78}$As, but the localization of exciton can enhance this value drastically, for example  $\Delta$ in GaAs is  enhanced by more than two orders of magnitude  for excitons confined in a quantum dot\cite{jmf99}. Further  experiments  are required to corroborate this scenerio.

In conclusion, we have observed  many-body effects  like BGN and  FES-like features  in moderately heavy doped  Al$_x$Ga$_{1-x}$As by PL spectroscopy.   FES arises due to  multiple
scattering of electrons near Fermi-edge by the photogenerated holes
localized by alloy fluctuation in Al$_x$Ga$_{1-x}$As.  A  two-level-based model has been proposed to  can explain the nonmonotonic temperature and excitation intensity dependence of FES related transition in PL spectra.  The present results indicate two directions for future research. More experimental and theoretical are  required for further ascertainment of (i) the   nonmonotonic temperature and excitation intensity dependence of FES and (ii) the effect of hole    localization on many-body effects leading to  the formation of FES.

\section*{Figure Captions}

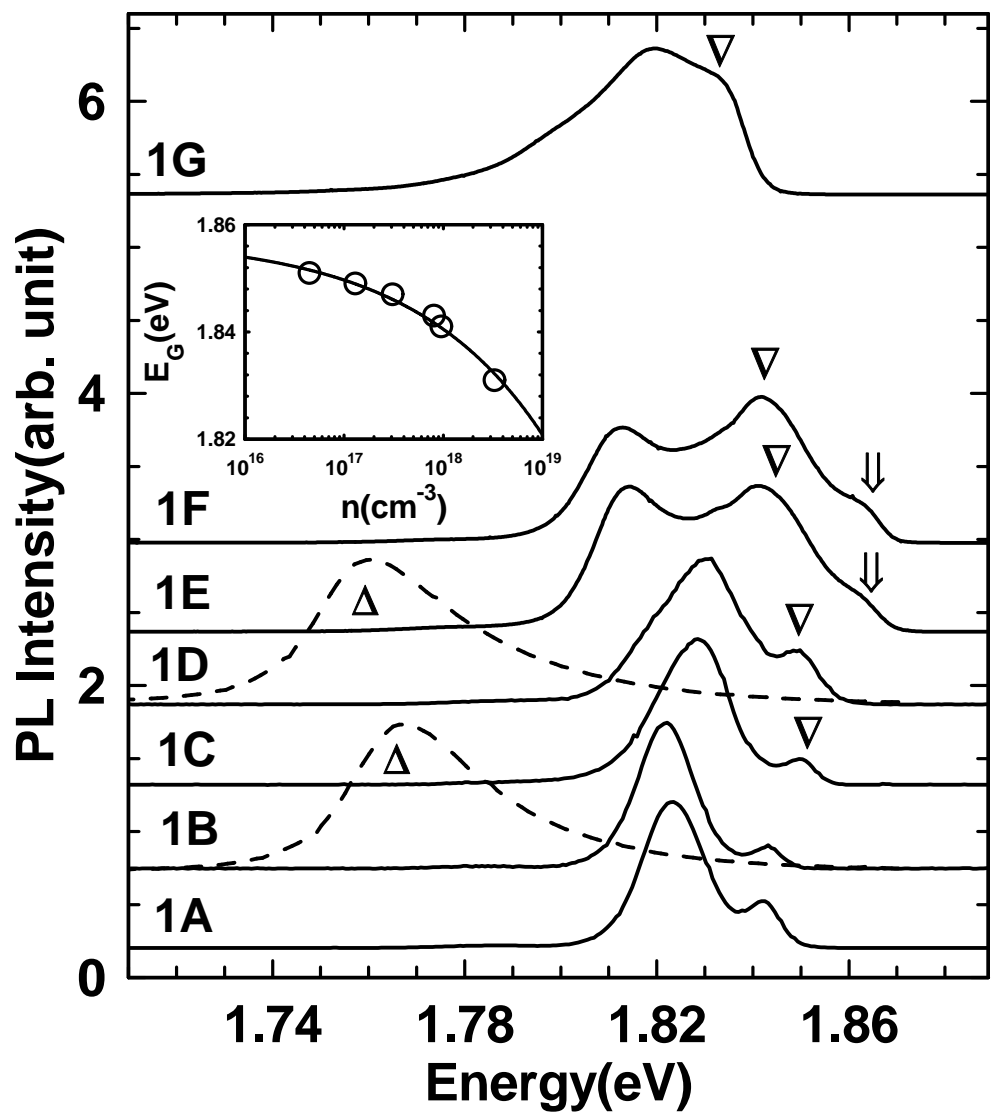
\noindent Fig.1. Evolution of PL spectra of Al$_{0.22}$Ga$_{0.78}$As as a function of  electron concentration.  Solid lines(for samples 1A, 1B, 1C, 1D, 1E, 1F, 1G)  represent PL spectra at  T=1.8K and dashed lines(for samples 1B and  1D) represent PL spectra at T=300K. Triangles and inverted triangles  indicate the spectral position of band-to-band transition at T=300K and 1.8K, respectively.      Inverted  arrows($\Downarrow$)  indicate the spectral position of the
FES-like feature in PL spectra. Inset shows the reduction of band gap as a function of logarithm of electron concentration.

\vspace{0.1in}

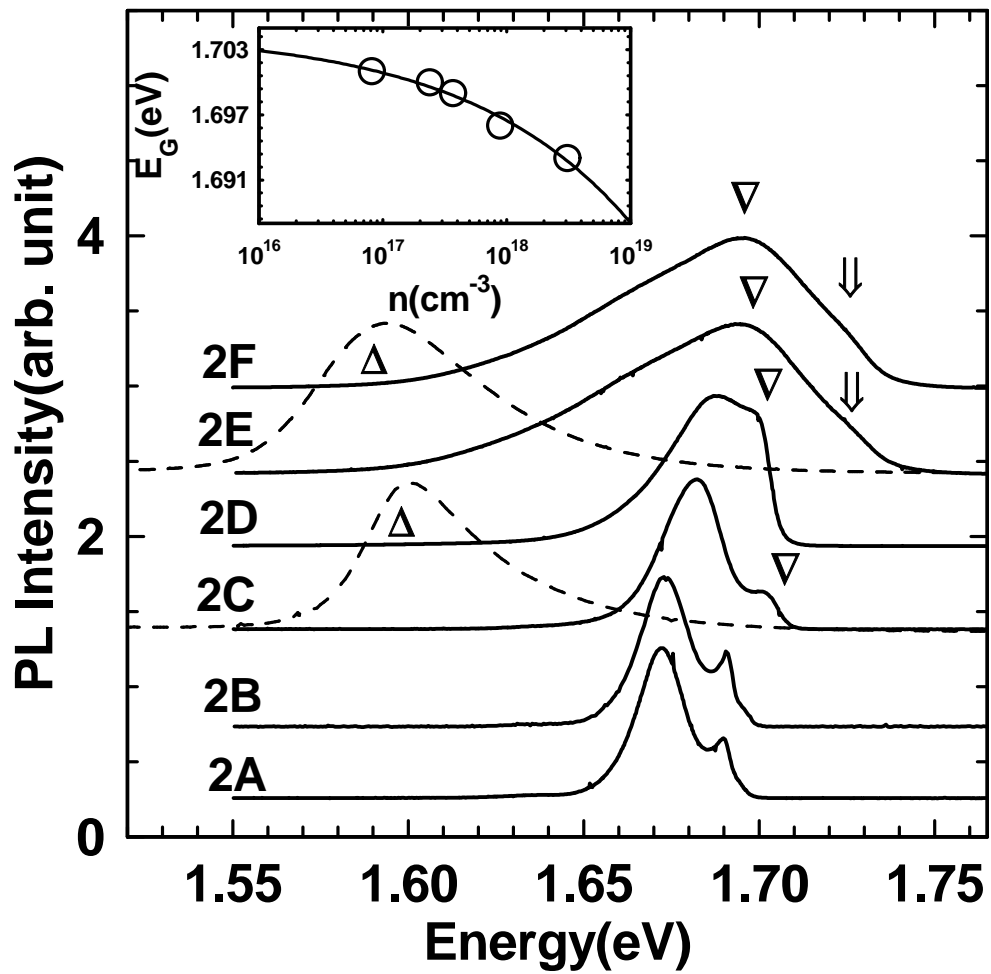
\noindent Fig.2. Evolution of PL spectra of Al$_{0.11}$Ga$_{0.89}$As as a function of  electron concentration.  Solid lines(for samples 2A, 2B, 2C, 2D, 2E, 2F)  represent PL spectra at  T=1.8K and dashed lines(for samples 2C and  2E) represent PL spectra at T=300K. Triangles and inverted triangles  indicate the spectral position of band-to-band transition at T=300K and 1.8K, respectively.      Inverted  arrows($\Downarrow$)  indicate the spectral position of the
FES-like feature in PL spectra. Inset shows the reduction of band gap as a function of logarithm of electron concentration.

\vspace{0.1in}

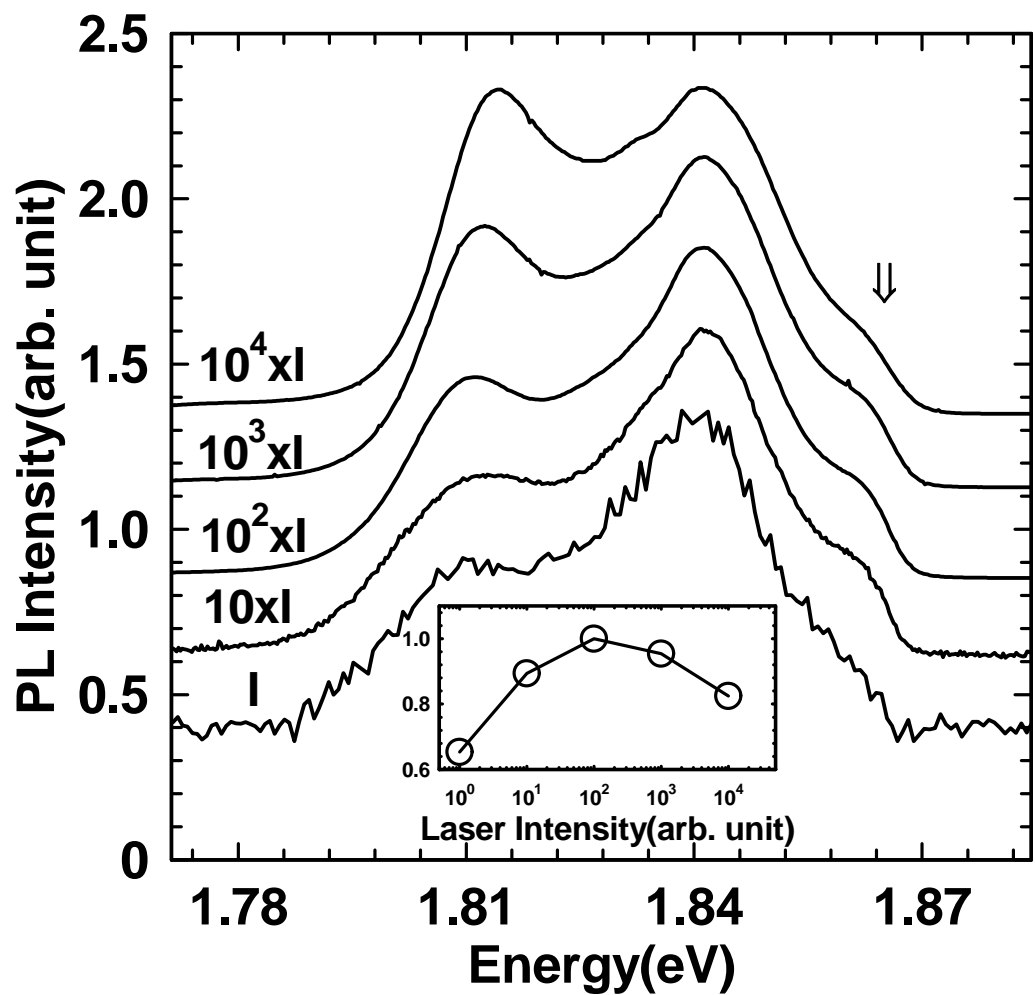
\noindent Fig.3. Excitation-intensity dependence of photoluminescence
spectra of Al$_{0.22}$Ga$_{0.78}$As with 9.6$\times$10$^{17}$cm$^{-3}$(sample 1F)
electron concentration  at 1.8K.  Arrows($\Downarrow$)  indicate  the spectral position of the FES-like feature in PL spectra. Inset shows the PL intensity of  FES-like feature as a function of logarithm of excitation intensity(I). Solid line is guide for the eyes.

\vspace{0.1in}

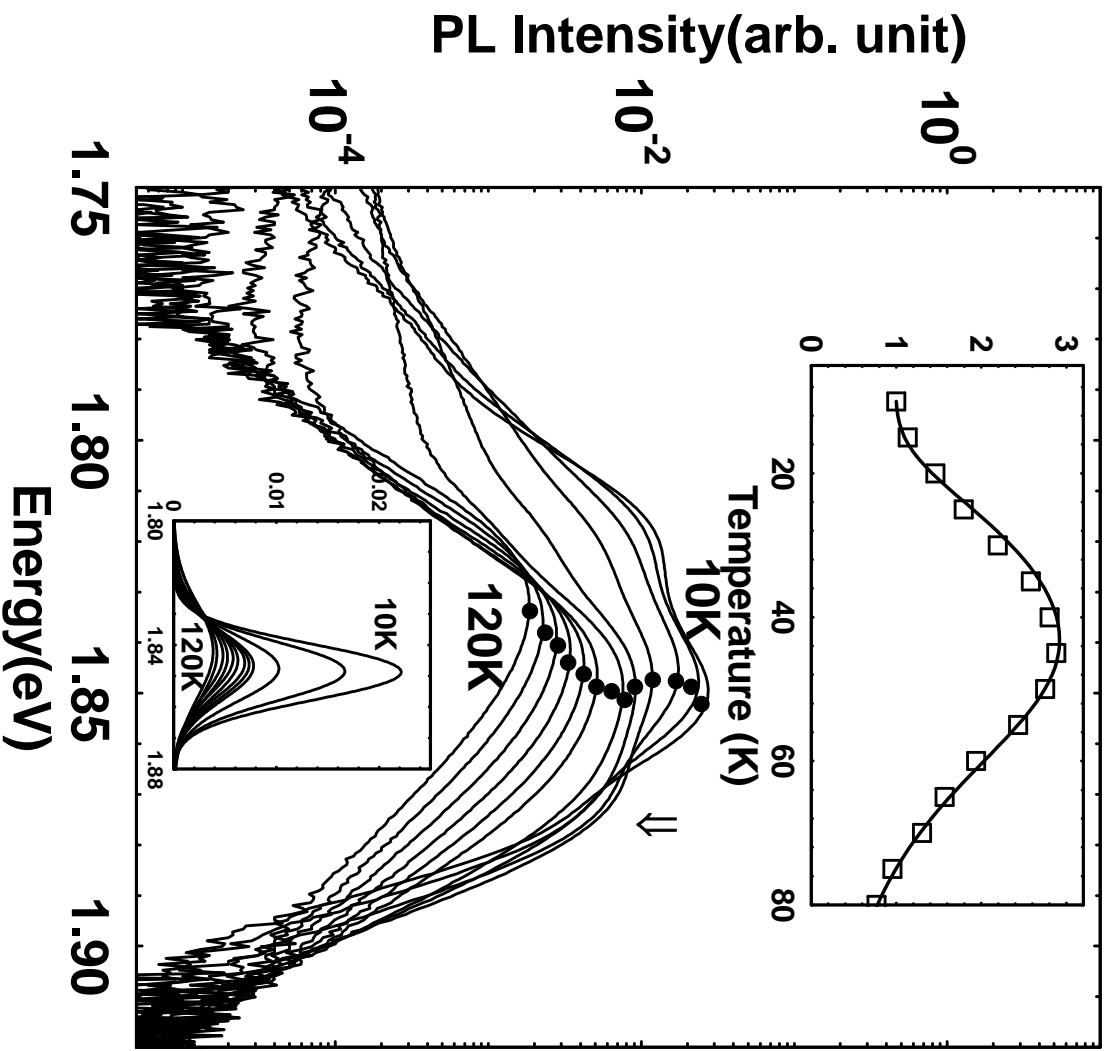
\noindent Fig.4. Temperature dependence  PL spectra of
Al$_{0.22}$Ga$_{0.78}$As with 9.6$\times$10$^{17}$ cm$^{-3}$(sample 1F) electron
concentration.  The emission peak due to band-to-band transition shows an anomalous S-shaped shift with increasing temperature(shown as solid circles) due  to the nonmonotonic temperature dependence of PL intensity of high energy FES-like peak, shown in upper inset. Lower inset shows the deconvoluted BB peak, which shows usual monotonic  redshift with increasing temperature.

\end{document}